\documentstyle[aps,preprint,tighten,eqsecnum]{revtex}
\def\beq{\begin{equation}}
\def\eeq{\end{equation}}
\def\beqn{\begin{eqnarray}}
\def\eeqn{\end{eqnarray}}

 \def\lsim{\mathrel{\rlap{\lower4pt\hbox{\hskip1pt$\sim$}}
    \raise1pt\hbox{$<$}}}
\def\slash#1{\setbox0=\hbox{$#1$}#1\hskip-\wd0\hbox to\wd0{\hss\sl/\/\hss}}
\begin{document}
\vspace{20pt}
\title{  DYNAMICAL GENERATION OF LIGHT FERMIONS}
\vspace{25pt}
\author{PANKAJ JAIN \footnote {e-mail: pkjain@iitk.ernet.in}}
 \address{Physics Department, I.I.T., Kanpur, India 208016}
 \maketitle
\vspace{35pt}
\begin{abstract}
We show that composite fermions with masses much smaller than the scale of
confinement arise naturally in certain models which admit dynamical
breakdown of chiral symmetry.
 The models are such that to leading order some of the fermions
remain massless but pick up small dynamical masses at subleading order.
\end{abstract}

\bigskip
\noindent
{\bf 1. Introduction}

\bigskip
The existence of fermions with masses much smaller than the scale of
electroweak symmetry breaking might be the result of an approximate chiral
symmetry of the underlying model. There exist several composite models
\cite{comp1}
based on this idea, but in most cases in the absence of fundamental scalars
some of the fermions remain exactly massless
and remaining fermions pick up masses  of the order of the scale of confinement
of the underlying strong dynamics, which may be equal to or larger than the
scale of
electro-weak symmetry breaking. Generation of nonvanishing masses for light
fermions in a dynamical framework has proven to be a very difficult problem
in all of the popular scenarios including technicolor \cite{techni},
top condensate \cite{top} as well as models in which fermions and/or
electroweak bosons are composite \cite{comp1}.
 In the present paper we display some situations
in which some of the fermions dynamically acquire
very small but non zero masses.

\bigskip
\noindent
{\bf 2. Light fermions in a large N chiral model}

\bigskip
Dynamical light fermions can arise if the fermion representation is such
that to leading order fermion condensate is prevented from forming. We have
in mind some nonabelian gauge group and by leading order we mean leading order
in either the loop expansion or the $1/N$ expansion \cite{largen}
 or a small gauge
coupling parameter. In the present section we will confine ourselves to the
cases in which $1/N$ is the small parameter and will discuss
the generalizations
in the later sections. The possibility of a systematic
loop expansion within a dynamical framework is very interesting although there
is no evidence that this is a reliable expansion scheme.

To give a simple example
we consider a SU(N)$_1\times$SU(N)$_2\times$SU(2)$_1\times$SU(2)$_2$
 model with N large. The small parameter
in this example is $1/N$.
We introduce
fermions in following representation of the
SU(N)$_1\times$SU(N)$_2\times$SU(2)$_1\times$SU(2)$_2$ gauge group.
$$a_L\rightarrow (N,1,1,1),\ \ \ \ a_R\rightarrow (1,N,1,1)$$
$$b_L\rightarrow (N,1,1,1),\ \ \ \ b_R\rightarrow (1,N,1,1)$$
$$c_L\rightarrow (N^*,1,2,1),\ \ \ \ c_R\rightarrow (1,N^*,1,2)$$
The representation has been chosen such that the interaction is free of
anomalies. We have not specified the transformation of these fermion
representations under electroweak and color interactions. Generalization
of the present model to include electroweak and color interactions will
be discussed in a separate publication.
The two SU(2) groups are assumed to be strong which will prevent the
condensation of the fermions $a$ or $b$ with the fermion $c$.
The strong SU(2) groups continue to have asymptotic freedom as long as $N<11$.
 As we discuss later
this model is expected to be confining because of its non-abelian nature and
the
gauge symmetry breaking is small.
We assume the scale of confinement of the two SU(N) groups
to be $\Lambda_{\rm conf}$.

The reason we are attracted to such SU(N)$_1\times$SU(N)$_2$ type models is
that the scale of chiral
symmetry breaking will be considerably suppressed compared to the scale of
confinement. Because of the presence
of the strong SU(2) interactions, the fermion $a_L$ can only condense with
either $a_R$ or $b_R$. We assume that it condenses only with one of these two
right handed particles and the horizontal symmetry which rotates $a$ into $b$
remains unbroken. Based on our experience with QCD we expect this to be true.
In any case we can always assign different electric charges to these two
fermions to assure that $a_L$ condenses only with $a_R$.
To analyze the pattern of mass generation we note that
 all possible diagrams that can convert a left
handed fermion to a right handed fermion contain atleast one internal fermion
loop and are therefore suppressed by one power of N.
All of these graphs
will therefore vanish as $N\rightarrow \infty$ and chiral symmetry will
remain unbroken. In arriving at this conclusion we have assumed that
the dynamically generated fermion mass decreases as the value of the
effective coupling decreases. Based on model calculations this is generally
expected to be the case for QCD type vectorial theories. For example,
the one loop analysis of Schwinger-Dyson shows that as $\alpha_s$, treated as
a constant, decreases
the dynamically generated mass also decreases and eventually vanishes as the
coupling goes below a critical value \cite{dsb}. If instead a renormalization
group improved expression is used for the effective coupling, such that the
coupling continues to rise with decrease in momentum, then there is
no critical point but the dynamical mass continues to decrease
with the decrease
in the value of effective coupling at some scale.
We will assume this also to be the case for the present model.
The Schwinger-Dyson (SD)
equation including only the leading order contribution
 in the loop expansion of CJT effective action \cite{cjt},
which can lead to chiral symmetry breaking, is shown is Fig. 1.

To see how the $1/N$ suppression
factor effects the scale of chiral symmetry breaking we use a simple model
for the nonabelian gauge coupling. We assume that the coupling has the form
$$\alpha \approx 1/{\rm log}\bigg(q^2/\Lambda^2_{\rm conf} + 1\bigg)$$
The reason for this choice \cite{richardson} is simply that it interpolates
between the
correct asymptotic behavior and the popular infrared behavior \cite{infra}
since it
leads to $1/q^4$ momentum dependence for the hypergluon propagator. We take
the scale at which this coupling becomes equal to 1 to be the scale of
chiral symmetry breaking. For the chosen behavior the coupling becomes
equal to 1 at $q^2/\Lambda^2_{\rm conf} = 1.7$. This is roughly the
scale of chiral
symmetry breaking if the theory is vectorial. However in the present case
the effective coupling is a factor of $N$ smaller. To get the
scale of chiral symmetry breaking in this case we set $\alpha/N=1$ to get
$q^2/\Lambda^2_{\rm conf} = {\rm exp}(1/N) -1$. For $N$ of the order of 10
this yields
$q^2/\Lambda^2_{\rm conf} = 0.1$, which shows a significant suppression factor.
This model calculation atleast shows that for N large enough the scale of
chiral symmetry is much smaller than the scale of confinement. We point out
that if N is arbitrarily large but not infinite than in the model
discussed above chiral symmetry breaking will take place. It is not clear,
however, if this
is true in reality because of our lack of knowledge of the behavior
of nonabelian theories at low energies and the effective coupling may
not increase monotonically with decrease in momentum.
 In any event there may exist a
large range of values of N for which the chiral symmetry still takes place.
We assume this to be the case.

  This theory will behave very
different from QCD in which case the scale of confinement is roughly the
same as the scale of chiral symmetry breaking. In particular, in the present
case to leading order we may simply ignore chiral symmetry breaking. Indeed
in the  limit as N$\rightarrow\infty$, there is no breakdown of chiral
symmetry.
 This implies that to leading order all fermions will be massless.
This conclusion holds not only for the elementary confined fermions but also
for the composite fermions which are bound states of the type
N$\times$N$\times$N$\times$..... . In the left-right theory under consideration
there will infact exist several such states, the simplest
one being made of N
left handed or N right handed fermions. More complicated composite fermions
may also be formed by including one or more hypergluons along with the
N fermions. In the infinite N limit these are the only type of fermions
allowed.
 One cannot, for example, have a bound state containing some left handed and
 some right handed fermions since there is no binding between left and
 right handed fermions in this limit.

We next consider the finite N corrections which will link the two
composite fermions discussed above. In order to convert a left handed composite
fermion to a right handed composite fermion we need to convert all the N
fundamental left handed fermions into right handed fermions. This coupling of
left and right handed composite fermions is shown in figure 2. Coversion of
each of these fundamental left handed fermion into a right handed fermion is
suppressed by one power of the ratio of the scale of chiral symmetry breaking
and the confinement scale. We call this suppression factor $\xi$. The
conversion of N elementary left handed fermions into right handed fermions
will therefore be suppressed
by $\xi^N$. This shows that the mass of these composite particles will be
extremely small. The logic used above to get this suppression factor is
very different from the usual intuition one has about bound states. However
even the usual intuition applied to the present case shows that the suppression
factor has to be very large. We are forming very tightly bound states of
fermions which have mass much smaller than the scale of confinement. The
binding
energy is necessarily very large, and results in very small bound state mass.
This argument, however, does not tell us whether the lightest state
is a fermion or a boson. The systematic large N expansion gives a very
good indication that the fermion has to be the lightest state.
The boson states,
which have a group theoretic structure N$\times \bar{\rm N}$,
are much heavier since they do not have the suppression factor $\xi^N$. The
mass of these states in the large N limit is independent of N and therefore
these states will have masses much smaller than the confinement scale but
not as small as the fermion masses.

We note that because of finite N corrections the gauge symmetry
SU(N)$_1\times$SU(N)$_2$ will be broken to SU(N)$_{hc}$, where $hc$ stands
for hypercolor. However since this breaking is subleading the massive gauge
particles will also be very light. Furthermore since the breaking is
negligible to leading order, the theory is confining and all the physical
states must be singlets under both SU(N)$_1$ and under SU(N)$_2$.

The model described in this paper naturally generates very tightly bound
composite fermions
which are much lighter than the scale of confinement. Generalizations of
this model to include color and electroweak interactions are currently
under consideration and will be described in a separate publication.

\bigskip
\noindent
{\bf 3. Acknowledgement}

\bigskip
  I would like to thank Chris Hill,
 Doug McKay, Herman Munczek, John Ralston, Joseph
  Schechter and
especially Kimball Milton for useful discussions. This work was
supported in part  by the U.S. DOE contract
no. DE-FG05-91ER-40636.

\vfill
\eject

\end{document}